\documentstyle[aps,preprint,tighten,epsfig,amstext,rotating]{revtex}

\newcommand{\psibar}{\bar\psi}
\newcommand{\chibar}{\bar\chi}
\newcommand{\bra}{\langle}
\newcommand{\ket}{\rangle}
\newcommand{\pbp}{\langle\bar\psi\psi\rangle}
\newcommand{\cbc}{\langle\bar\chi\chi\rangle}
\newcommand{\U}[1]{\mathrm{U}(#1)}

\newcommand{\One}{1\kern-4.5pt1}
\newcommand{\half}{\frac{1}{2}}
\newcommand{\Tr}{\mathrm{Tr}}
\newcommand{\cc}[1]{#1}
\newlength{\colw}
\newcommand{\m}{\hphantom{$-$}}

\begin{document}

\preprint{DESY 00-175}

\title{Lattice simulations of QCD-like theories at non-zero density}

\author{
Jonivar Skullerud}
\address{
DESY Theory Group, Notkestra{\ss}e 85, D--22603 Hamburg, Germany}

\maketitle

\begin{abstract}
One way of avoiding the complex action problem in lattice QCD at
non-zero density is to simulate QCD-like theories with a real action,
such as two-colour QCD.  The symmetries of two-colour QCD with quarks
in the fundamental and in the adjoint representation are described,
and the status of lattice simulations is reviewed, with particular
emphasis on comparison with predictions from chiral perturbation
theory.  Finally, we discuss how the lessons from two-colour QCD may
be carried over to physical QCD.
\end{abstract}

\section{Introduction}
\label{sec:intro}

Recently, there has been a considerable interest in QCD at non-zero
chemical potential, after a number of model studies have indicated a
rich phase structure \cite{Alford:1998mk,Alford:1997zt,Rapp:1998zu}
(for a review, see \cite{Rajagopal:2000wf}).  Clearly, it would be
desirable if these predictions could be tested by first-principles,
non-perturbative studies, e.g.\ lattice QCD.  Unfortunately, lattice
simulations of QCD at non-zero baryon density using standard methods
are in practice impossible because the action (and the fermion
determinant) becomes complex once the chemical potential is
introduced, causing importance sampling to fail.  It is possible to
split the determinant into a modulus and a phase, simulating with the
modulus of the determinant as the measure and reweighting the
observables with the phase,
\begin{equation}
\bra O\ket = \frac{\bra\bra O\,\mbox{arg(det} M)\ket\ket}
                    {\bra\bra \mbox{arg(det} M)\ket\ket},
\label{eq:obs}
\end{equation}
where $\bra\bra\ldots\ket\ket$ denotes the expectation value with
respect to the positive real measure.  However, the denominator in
(\ref{eq:obs}) is effectively the ratio of the partition functions of
two different theories: the true theory and one with a positive real
measure.  This should scale as $\exp(-\Delta F)$, where $\Delta F$ is
the difference in free energy between the two theories.  Since the
free energy is an extensive quantity, the computational effort
required to obtain a reliable sample rises exponentially with the
volume.

A number of approaches have been tried to overcome this problem.  In
the Hamiltonian formalism, the problem does not arise.  Analytical
results have been obtained in the strong coupling limit
\cite{Gregory:1999pm,Luo:2000xi}, but so far no method for numerical
simulations exists.

With an imaginary chemical potential \cite{Alford:1998sd}, the action
becomes real and positive, so simulations are straightforward.  The
problem is whether an analytical continuation to real $\mu$ is
possible.  It works at high temperature
\cite{Lombardo:1999cz,Hart:2000ef}, where also other approaches may be
successfully employed \cite{Hart:2000ef}, but does not seem to be
possible at zero or low temperatures.  Imaginary chemical potential
can also be used to formulate a quenched limit of QCD in the
background of a non-zero number of static quarks \cite{Engels:1999tz}.

Cluster algorithms \cite{Chandrasekharan:2000ew} may provide a way of
eliminating the sign problem by summing analytically over
configurations in a cluster in such a way that the contribution to the
partition function from each cluster is always positive definite.  So
far, these methods have been applied to a number of spin models;
however, an application to QCD has yet to be found.

Finally, the sign problem may be avoided by simulating theories which
resemble QCD, but have a real action even at non-zero chemical
potential.  One such theory is QCD at non-zero isospin density
\cite{Son:2000xc,Hands:2000hi}, which is of intrinsic interest because
it corresponds to part of the phase diagram for asymmetric nuclear
matter.  Another class of theories encompasses two-colour QCD with
fermions in the fundamental representation, as well as QCD with
adjoint fermions, for any number of colours.  The remainder of this
review will focus on what can be learnt from lattice simulations of
these theories.

\section{Theories with real action}
\label{sec:realaction}

The chemical potential $\mu$ is introduced on the lattice by
multiplying the forward timelike links by $e^\mu$ and the backward
timelike links by $e^{-\mu}$ \cite{Hasenfratz:1983ba}.  It can be
shown \cite{Kogut:2000ek,Hands:2000ei} that the determinant $\det{M}$
of the fermion matrix $M$ in two-colour QCD is real, even for non-zero
$\mu$, both in the continuum and on the lattice.  However, it is only
possible to demonstrate that it is positive \cite{Hands:2000ei} in the
cases of continuum or Wilson adjoint fermions and staggered
fundamental fermions.  Indeed, we will see in section
\ref{sec:adjoint} that in the case of staggered adjoint fermions there
are configurations with a negative determinant, leading to a sign
problem at large $\mu$.

\subsection{Symmetry breaking pattern}
\label{sec:symmbreaking}

In the chiral limit, the action for two-colour QCD with $N$ flavours
has a $\U{N}_L\otimes\U{N}_R$ symmetry, which for staggered fermions
is manifest as independent $\U{N}$ symmetries for the even and odd
sites.  At $\mu=0$ this enlarges to a $\U{2N}$ symmetry.  This can be
seen most easily by introducing new fields,
\begin{equation}
\bar{X}_e = (\chibar_e,-\chi^T_e\tau_2) \qquad X_o = 
\left(\begin{array}{c}
\chi_o\\-\tau_2\bar\chi_o^T
      \end{array} \right) \\
\end{equation}
for (staggered) fundamental quarks, and
\begin{equation}
\bar{X}_e = (\chibar_e,\chi^T_e) \qquad X_o = 
\left(\begin{array}{c}
\chi_o\\\bar\chi_o^T
      \end{array}\right)
\end{equation}
for adjoint quarks.  The action can then be written as
\begin{eqnarray}
S={1\over2}\sum_{x_e, \nu}\eta_\nu(x)
\biggl[\bar X_e(x)&\left(\matrix{e^{\mu\delta_{\nu,0}}&0\cr
             0 &e^{-\mu\delta_{\nu,0}}\cr}\right)&
                           U_\nu(x)X_o(x+\hat\nu) \, -  \\
\bar X_e(x)&\left(\matrix{e^{-\mu\delta_{\nu,0}}&0\cr
             0 &e^{\mu\delta_{\nu,0}}\cr}\right)&
U_\nu^\dagger(x-\hat\nu)X_o(x-\hat\nu)\biggr] \nonumber
\end{eqnarray}
where $x_e$ denotes the even sites. In the continuum, the equivalent
fields are
\begin{equation}
\text{fundamental:} \quad 
\Psi = \left(\begin{array}{c}\psi_L\\
	 \sigma_2\tau_2\psi_R^{*}\end{array}\right)
\qquad
\text{adjoint:} \quad
\Psi = \left(\begin{array}{c}\psi_L\\ \sigma_2\psi_R^{*}\end{array}\right)
\end{equation}
which gives the continuum lagrangian
\begin{equation}
{\mathcal{L}} = i \Psi^\dagger \sigma_\nu(D_\nu-\mu B_\nu)\Psi \,
\qquad B_\nu = \left(\begin{array}{cc} 1 & 0 \\ 0 & -1
\end{array}\right) \delta_{\nu0} \, .
\end{equation}
The explicity chiral symmetry breaking term in the Wilson fermion
action means that there is no equivalent enlarged symmetry for Wilson
fermions; however, new fields may be introduced analogously to the
continuum case, and the enlarged symmetries will be broken by 
${\cal O}(a)$ terms in the action.

The chiral condensate can be written in terms of the new fields,
\begin{equation}
\bar\chi\chi=
\bar X_e\!\left(\begin{array}{cc}0&\One\\
		\pm\One&0\end{array}\right)\!{T\over2}\bar X_e^{tr}+
X_o^{tr}\!\left(\begin{array}{cc}0&\One\\
		\pm\One&0\end{array}\right)\!{T\over2}X_o
\label{eq:condf}
\end{equation}
for staggered fermions, and
\begin{equation}
\psibar\psi = \Psi^T\sigma_2\frac{T}{2}
\left(\begin{array}{cc}0&-\One\\\pm\One&0\end{array}\right)
\Psi + {\rm h.c.}
\label{eq:cond-cont}
\end{equation}
in the continuum and for Wilson fermions.  In both cases, the $+$ sign
is for fundamental fermions and the $-$ sign for adjoint, while $T$ is
$\tau_2$ for fundamental fermions and 1 for adjoint.  A nonzero chiral
condensate thereby breaks down the $\U{2N}$ symmetry to O($2N$) for
fundamental fermions and Sp($2N$) for adjoint fermions, giving rise to
$N(2N+1)$ and $N(2N-1)$ Goldstone modes respectively.  Of these, there
will be $N^2$ mesonic states, while the remaining $N(N\pm1)$ will be
diquarks.  In the continuum, and for Wilson fermions, the pattern will
be the opposite (modulo the 1 mode destroyed by the axial anomaly in
the continuum), but for 1 flavour of fundamental quarks, there is no
chiral symmetry in the first place.  From this we see that in the case
of $N=1$ adjoint staggered fermions, and only in this case, are there
no diquark Goldstone modes.

For $m\neq0$, all the pseudo-Goldstone modes remain degenerate, with
masses $m_\pi\propto\sqrt{m}$.  As the chemical potential $\mu$
increases, the ground state will begin to be populated with baryonic
matter.  The transition to a ground state containing matter occurs
when $\mu=\mu_o\simeq m_b/n_q$, where $m_b$ is the mass of the
lightest baryon, and this baryon contains $n_q$ quarks.  At this
point, the baryon number density $n$ becomes non-zero, where $n$ is
given by
\begin{eqnarray}
\label{eq:n-wilso}
n & =  \half\big\bra&\psibar(x)e^\mu(\gamma_0-1)U_0(x)\psi(x+\hat0) \\
 & & +
 \psibar(x+\hat0)e^{-\mu}(\gamma_0+1)U_0^\dagger(x)\psi(x)\big\ket
\nonumber
\end{eqnarray}
for Wilson fermions, and
\begin{equation}
n=\half\big\bra\bar\chi(x)\eta_0(x)[e^\mu U_0(x)\chi(x+\hat0)
 +  e^{-\mu}U_0^\dagger(x-\hat0)\chi(x-\hat0)]\big\ket. 
\label{eq:n}
\end{equation}
for staggered fermions.
Where there are diquark Goldstone modes, those states will be the
lightest baryons in the spectrum. This means that for most variants of
two-colour QCD we expect $\mu_o\simeq m_\pi/2$, in contrast to the
much larger value $m_N/3$ expected in real (three-colour) QCD.  The
exception is two-colour QCD with one flavour of adjoint staggered
quarks.

In the limit of small $m$ and $\mu$, the behaviour of $\pbp$, the
diquark condensate $\bra\psi\psi\ket$, and $n$
as functions of $m$ and $\mu$ can be calculated in chiral perturbation
theory.  If we define the rescaled variables
\begin{equation}
x=\frac{2\mu}{m_{\pi0}} \,, \quad 
y=\frac{\pbp}{\pbp_0} \,, \quad z=\frac{\bra\psi\psi\ket}{\pbp_0}\,, \quad 
\tilde n=\frac{m_{\pi0}n}{8m\pbp_0} \, ,
\label{eq:rescale}
\end{equation}
where the 0 subscript denotes values at $\mu=0$, the prediction from
$\chi$PT for the models with diquark Goldstone modes is
\cite{Kogut:2000ek}
\begin{equation}
y=\left\{\begin{array}{c}
	1\\
	1\over x^2
	 \end{array} \right.
\quad
z=\left\{\begin{array}{c}
	0\\\sqrt{1-\frac{1}{x^4}}
	 \end{array}\right.
\quad
\tilde n=\left\{\begin{array}{cl}
	0&;x\!<\!1\\
	{x\over4}\left(1-{1\over x^4}\right)&;x\!>\!1
	       \end{array}\right.
\label{eq:chipt}
\end{equation}

\subsection{Diquark condensation}
\label{sec:diquark}

At large chemical potential, the relevant degrees of freedom will be
quarks with momenta near the Fermi surface.  The attractive
quark--quark interaction will give rise to instability with respect to
condensation of diquark pairs at opposite sides of the Fermi surface.
In physical QCD, the diquark condensate cannot be a colour singlet, so
the gauge symmetry is spontaneously broken, giving rise to the
phenomenon of colour superconductivity.

In two-colour QCD, on the other hand, there may be the possibility of
gauge singlet diquarks, which will be energetically favoured compared
to non-singlet states.  Indeed, in the previous section we saw that
most variants of two-colour QCD have diquark Goldstone modes, which
will be the preferred channel for diquark condensation.  In the $N=1$
staggered adjoint model, this is not the case, and we do not know {\em
a priori} in which channel the condensation will occur.  We must
proceed by constructing possible operators which obey the Pauli
principle and making additional assumptions about locality, Lorentz
structure and gauge invariance \cite{Hands:2000ei}.  At least one of
the possible condensates constructed this way gives rise to a colour
superconding ground state.

The standard way of computing the diquark condensate on the lattice is
to introduce a diquark source term into the action
\cite{Morrison:1998ud}.  For two-colour QCD with fundamental staggered
quarks the action then becomes
\begin{eqnarray}
S_F & = & \sum_{x,y}\bar\chi(x)M_{xy}\chi(y) 
+ \sum_x\frac{j}{2}\left[\chi^T(x)\tau_2\chi(x)
  +\bar\chi(x)\tau_2\bar\chi^T(x)\right] \\
 & = & (\bar\chi,\chi^T)
 \left(\begin{array}{cc} j\tau_2 & \half M \\
  \half M & j\tau_2 \end{array}\right)
 \left(\begin{array}{c} \!\bar\chi^T\\ \!\!\chi\end{array}\right)
\equiv X^T{\mathcal{A}}[j]X \, .
\label{eq:action-diquark}
\end{eqnarray}
In this case, the partition function becomes proportional to the
Pfaffian $\text{Pf}{\mathcal{A}}[j]$.  The diquark
condensate $\bra\chi^T\tau_2\chi\ket$ may be evaluated by taking
\begin{equation}
\bra\chi^T\tau_2\chi\ket = \lim_{j\to0}\frac{1}{2V}
 \left\bra\Tr\left\{{\mathcal{A}}^{-1}
\left(\begin{array}{cc}\tau_2&0\\0&\tau_2\end{array}\right)\right\}\right\ket
\label{eq:diquark-j}
\end{equation}
An alternative approach \cite{Aloisio:2000nr} is to rewrite the
Pfaffian as
\begin{equation}
\mathrm{Pf}{\mathcal{A}}[j] = \mathrm{Pf}(B+j) = \pm\sqrt{\det(B+j)}
\label{eq:pfaff-expand}
\end{equation}
where
\begin{equation}
B = \left(\begin{array}{cc} 0 & \half M\tau_2\\ -\half M\tau_2 &
0\end{array}\right) \, .
\end{equation}
This can be expanded as a polynomial in $j$ by diagonalising $B^2$,
obviating the need to simulate at non-zero diquark source.  Since this
gives the Pfaffian at any $j$, it can also be used to determine the
diquark condensate using the probability distribution function
\cite{Azcoiti:1995dq}. 

\section{Simulations with fundamental quarks}
\label{sec:fund}

In the past year and a half, a number of groups have been performing
lattice simulations of two-colour QCD with fundamental staggered
fermions both at zero
\cite{Hands:1999zv,Aloisio:2000if,Aloisio:2000rb,Bittner:2000rf,Hands:2000hi}
and non-zero \cite{Alles:2000qi,Liu:2000in} temperature.  Also, one
group is performing simulations with Wilson fermions
\cite{Muroya:2000qp}.

\begin{figure}[p]
\begin{center}
\begin{sideways}
\epsfig{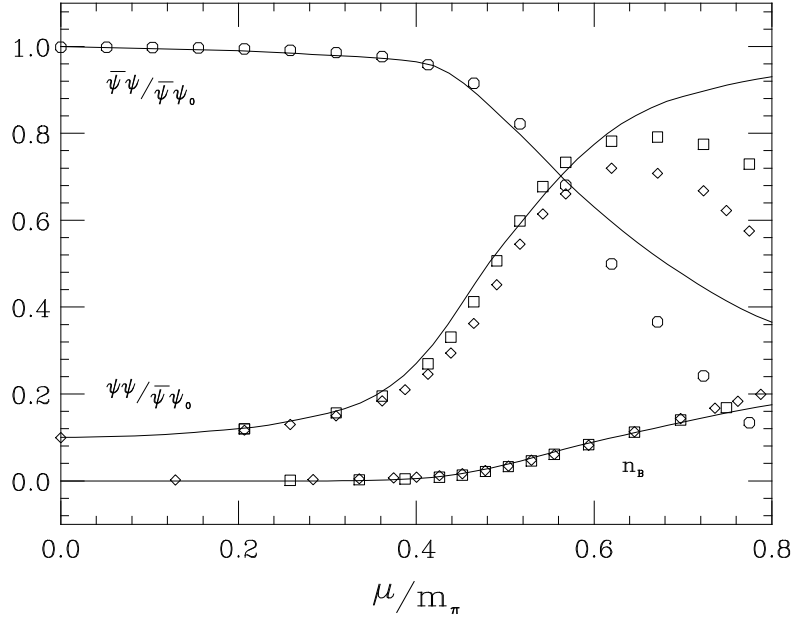}
\end{sideways}
\end{center}
\caption{Baryon density, chiral condensate and diquark condensate
vs. chemical potential, from \protect\cite{Aloisio:2000if}.  The solid
lines are the predictions of (\protect\ref{eq:chipt}).}
\label{fig:aloisio-eos}
\end{figure}

\begin{figure}[p]
\begin{center}
\begin{sideways}
\epsfig{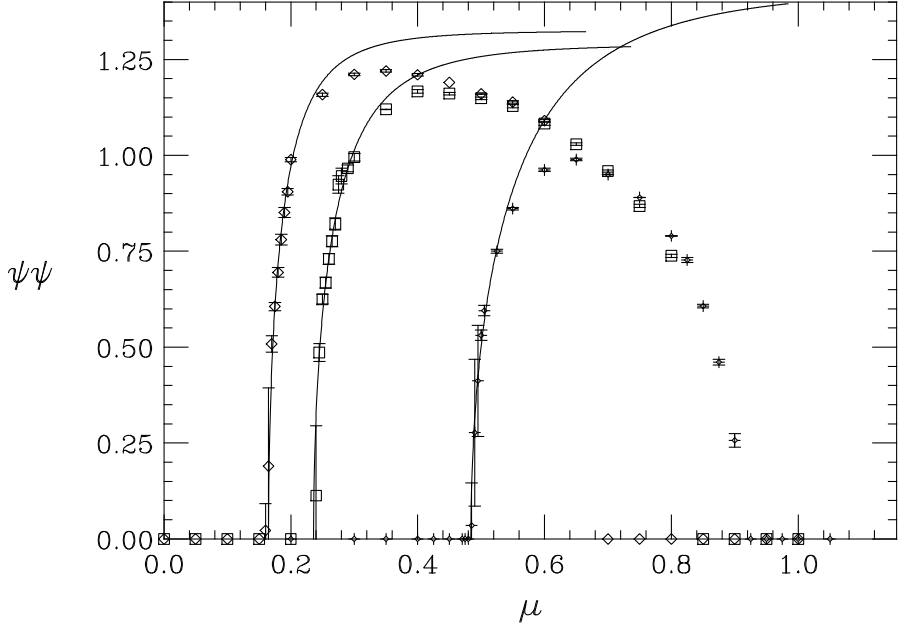}
\end{sideways}
\end{center}
\caption{Diquark condensate for a $6^4$ lattice, $N_f=1$, for
$m=0.025$ (diamonds), 0.05 (squares) and 0.2 (stars) at strong
coupling, from \protect\cite{Aloisio:2000rb}, with the predictions of
(\protect\ref{eq:chipt}).} 
\label{fig:aloisio-diq}
\end{figure}

Aloisio {\em et al.}\ \cite{Aloisio:2000if,Aloisio:2000rb} have
performed simulations in the strong coupling limit for a number of
quark masses, flavours and lattice volumes.
Fig.~\ref{fig:aloisio-eos} shows results for the chiral condensate,
the diquark condensate and the baryon number density, at $m=0.2$ and a
non-zero source $j=0.02$, for two different lattice sizes.  Also shown
are the $\chi$PT predictions from (\ref{eq:chipt}).
The agreement between the prediction and the numerical results is
quite striking, considering that this is far from the continuum
limit.  This suggests a weak $\beta$ dependence. 
Fig.~\ref{fig:aloisio-diq} shows the diquark condensate at zero
diquark source, for $N_f=1$ and a range of quark masses.  Again, we
see a very good agreement with the prediction (\ref{eq:chipt}).
At larger $\mu$, we see that the value of $\bra\psi\psi\ket$ drops,
going to zero at high $\mu$.  This second transition is due to lattice
artefacts connected with the saturation of lattice sites with
fermions.  In the infinite volume limit it is expected to disappear.

Simulations at non-zero $\mu$ with a diquark source have been
performed by Kogut and Sinclair \cite{Hands:2000hi}.  Results for the
diquark condensate and chiral condensate are shown in
fig.~\ref{fig:sinclair-cond}.  Again, we see the chiral condensate
dropping and the diquark condensate rising for $\mu\gtrsim m_\pi/2$,
in agreement with $\chi$PT.  We also see the same large-$\mu$
saturation behaviour for the diquark condensate as in
\cite{Aloisio:2000rb}. 
Fig.~\ref{fig:sinclair-mass} shows results for pion and scalar diquark
masses.  The scalar diquark mass falls roughly as $m_\pi-2\mu$ as
$\mu$ approaches $m_\pi/2$.  The pion mass remains constant up to
$\mu\approx m_\pi/2$, after which it falls to zero.  This is again in
accordance with the expectation from $\chi$PT.

The spectrum of the Dirac operator has been studied in some detail by
the Vienna group \cite{Bittner:2000rf} for staggered fermions and by
the Hiroshima group \cite{Muroya:2000qp} for Wilson fermions.  A
preliminary study of topology at non-zero temperature has also been
performed \cite{Alles:2000qi}.

\section{Simulations with adjoint quarks}
\label{sec:adjoint}

As indicated in section \ref{sec:symmbreaking}, two-colour QCD with
one flavour of adjoint staggered fermions has features which makes it
in some senses more `QCD-like' than other variants of two-colour QCD.
In particular, it has no diquark Goldstone modes, so we expect an
onset transition at a value of the chemical potential different from
$m_\pi/2$ --- possibly at $\mu=m_N/3$ where $m_N$ denotes the mass of
the lightest three-quark baryon (the `nucleon').  It also has a sign
problem.

\begin{figure}[tb]
\begin{center}
\epsfig{width=0.70\colw,file=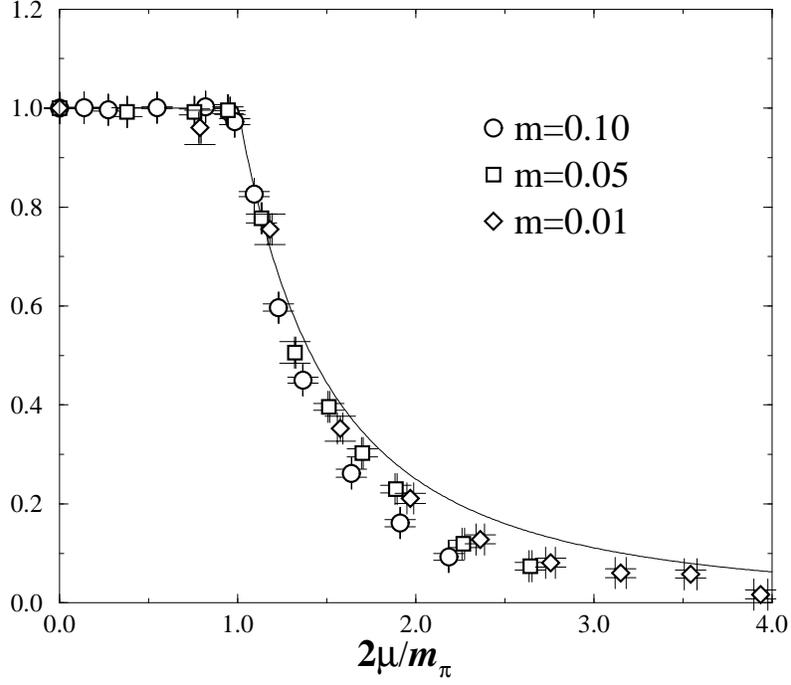}
\end{center}
\caption{Chiral condensate vs. chemical potential for one flavour of
adjoint staggered quarks,
using the rescaled variables of eq. (\ref{eq:rescale}); from
\protect\cite{Hands:2000yh}. 
\label{fig:unipbp}}
\end{figure}

\begin{figure}[tb]
\begin{center}
\epsfig{width=0.70\colw,file=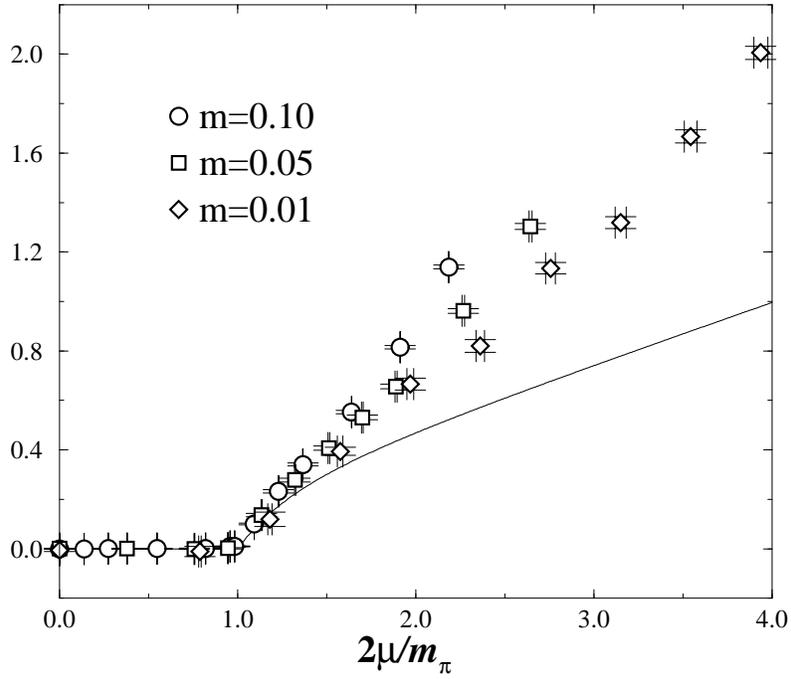}
\caption{Baryon density vs. chemical potential for one flavour of
adjoint staggered quarks, using the rescaled variables
of eq. (\ref{eq:rescale}); from \protect\cite{Hands:2000yh}.
\label{fig:uniden}}
\end{center}
\end{figure}

The theory has been simulated \cite{Hands:2000ei,Hands:2000yh} using
two different algorithms: Hybrid Monte Carlo, which is not able to
change the sign of the determinant, and therefore only simulates the
positive determinant sector of the theory, and the two-step
multibosonic algorithm \cite{TSMB}, which is able to take the sign
properly into account.  Figs~\ref{fig:unipbp} and \ref{fig:uniden}
show the results from HMC simulations for $y$ and $\tilde n$ of
(\ref{eq:rescale}) respectively, for a range of values for the quark
mass $m$ and chemical potential $\mu$.  Up to $x\sim1.5$, the data
collapse onto a universal curve, which agrees well with the
predictions of (\ref{eq:chipt}).  Even at larger $x$, the data for the
chiral condensate lie close to the $\chi$PT prediction.  At
$x\gtrsim2$, however, the data for different $m$ diverge, indicating
that $\chi$PT may be breaking down.  Results for the plaquette
\cite{Hands:2000yh} show a drop in its value for $\mu\geq m_\pi/2$,
presumably due to Pauli blocking, while the pion mass appears to agree
with the $\chi$PT prediction $m_\pi=2\mu$ at $x>2$.

The agreement between the predictions of $\chi$PT and these results
paradoxically enough presents a problem, since this model is not
supposed to contain any diquark Goldstone modes, and thus the $\chi$PT
predictions of (\ref{eq:chipt}) are not valid in this case.  In
particular, there should not be any onset transition at $\mu=m_\pi/2$.
The suspicion must be that this contradiction is due to the fact that
HMC does not change the sign of the determinant, and that it therefore
simulates the wrong theory --- a theory with conjugate quarks.

The simulation points for the TSMB algorithm were selected to focus on
the effect of the sign, with one point at $\mu=0$, one just past the
HMC onset transition, and one deeper into the dense region.  With this
algorithm, a reweighting factor $r$ and the sign of $\det{M}$ must be
determined for each configuration \cite{Montvay:1999ty}.  The
expectation value of an observable $O$ is then determined by the ratio
\begin{equation}
\bra O\ket={{\bra O\times r\times sign\ket}\over
                  {\bra r\times sign\ket}}.
\label{eq:Osign}
\end{equation}
The results for $\pbp$ and $n$, together 
with the corresponding HMC results, are summarised in Table
\ref{tab:tsmb}. For TSMB at $\mu\not=0$ we also include 
observables determined separately in each sign sector, defined by
$\bra O\ket_\pm=\bra O\times r\ket_\pm/\bra r\ket_\pm$.
At $\mu=0.0$ the two algorithms agree, as they should.  Also, the
results in the positive determinant sector for TSMB at larger $\mu$
agree with the HMC results.
However, the results for the negative determinant sector are
significantly different.  This difference has the effect of bringing
the average both for $\pbp$ and for $n$ back to values consistent with
the $\mu=0$ values.  This is an indication that at $\mu=0.36$ and
quite possibly also at $\mu=0.4$, the system is still in the vacuum
phase, which means that the onset transition in this model occurs at a
larger $\mu$ than for other variants of two-colour QCD.  This is
consistent with the symmetry-based
arguments of section \ref{sec:symmbreaking} that this model has no
baryonic Goldstone modes.

\section{Conclusions}

Substantial progress has been made recently in lattice simulations of
two-colour QCD at non-zero density, with both fundamental and adjoint
quarks.  The simulations with fundamental quarks nicely reproduce the
predictions from chiral perturbation theory for the chiral condensate,
diquark condensate, and baryon density, except at very large chemical
potentials.  Here, saturation effects are observed, especially for the
diquark condensate.  The meson and diquark spectrum is also being
analysed, with preliminary results for the pion and scalar diquark
masses again in rough agreement with chiral perturbation theory.

Two-colour QCD with one flavour of adjoint staggered quark is not
expected to have any diquark Goldstone modes, unlike all other
variants of two-colour QCD.  It also has a sign problem.  Simulations
of this model restricted to the sector with a positive fermion
determinant reproduce the predictions of chiral perturbation theory
for theories with diquark Goldstone modes, including an early onset
transition at $\mu\approx m_\pi/2$ --- although the breakdown of
$\chi$PT may be observed at larger $\mu$.  When configurations with
negative determinants are included, we find a strong correlation
between the sign and the value of observables.  This effect appears to
lead to a cancellation of the early onset transition.  This
observation may hold the key to understanding the problem of the
premature onset which has bedevilled previous attempts to simulate
physical QCD at non-zero density.

The presence of diquark modes which are degenerate with the pion means
that the study two-colour QCD is of limited usefulness when it comes
to studying directly the onset transition, hadron spectrum and diquark
condensation in physical QCD at non-zero density.  However, useful
experience may be gained by comparing the results of lattice
simulations with those of other methods which are also applicable to
physical QCD.  Of particular interest would be the study of
gluodynamics, where SU(2) and SU(3) are expected to exhibit similar
behaviour, even at non-zero $\mu$.  Thus it might be possible to cast
light on the deconfinement transition at high $\mu$ and low $T$ by
studying two-colour QCD.

\section*{Acknowledgements}
This work is supported by  the TMR network ``Finite temperature phase 
transitions in particle physics'', EU contract ERBFMRX--CT97--0122.

\setlength{\unitlength}{1.3cm} 
\begin{figure}
\begin{picture}(11.5,6)
\put(-0.5,0){\vbox{\epsfig{width=0.5\colw,file=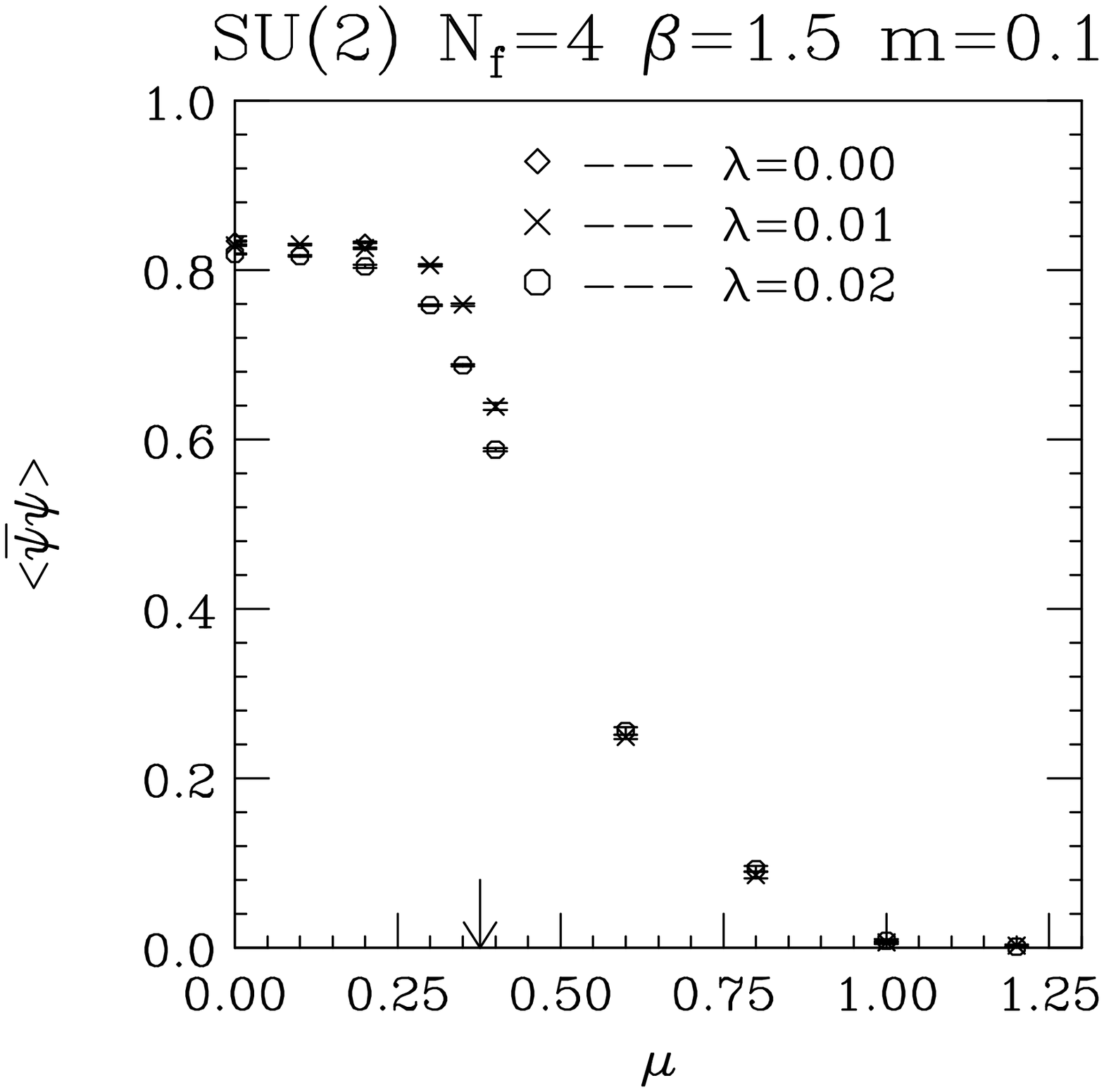}}}
\put(5.5,0){\vbox{\epsfig{width=0.5\colw,file=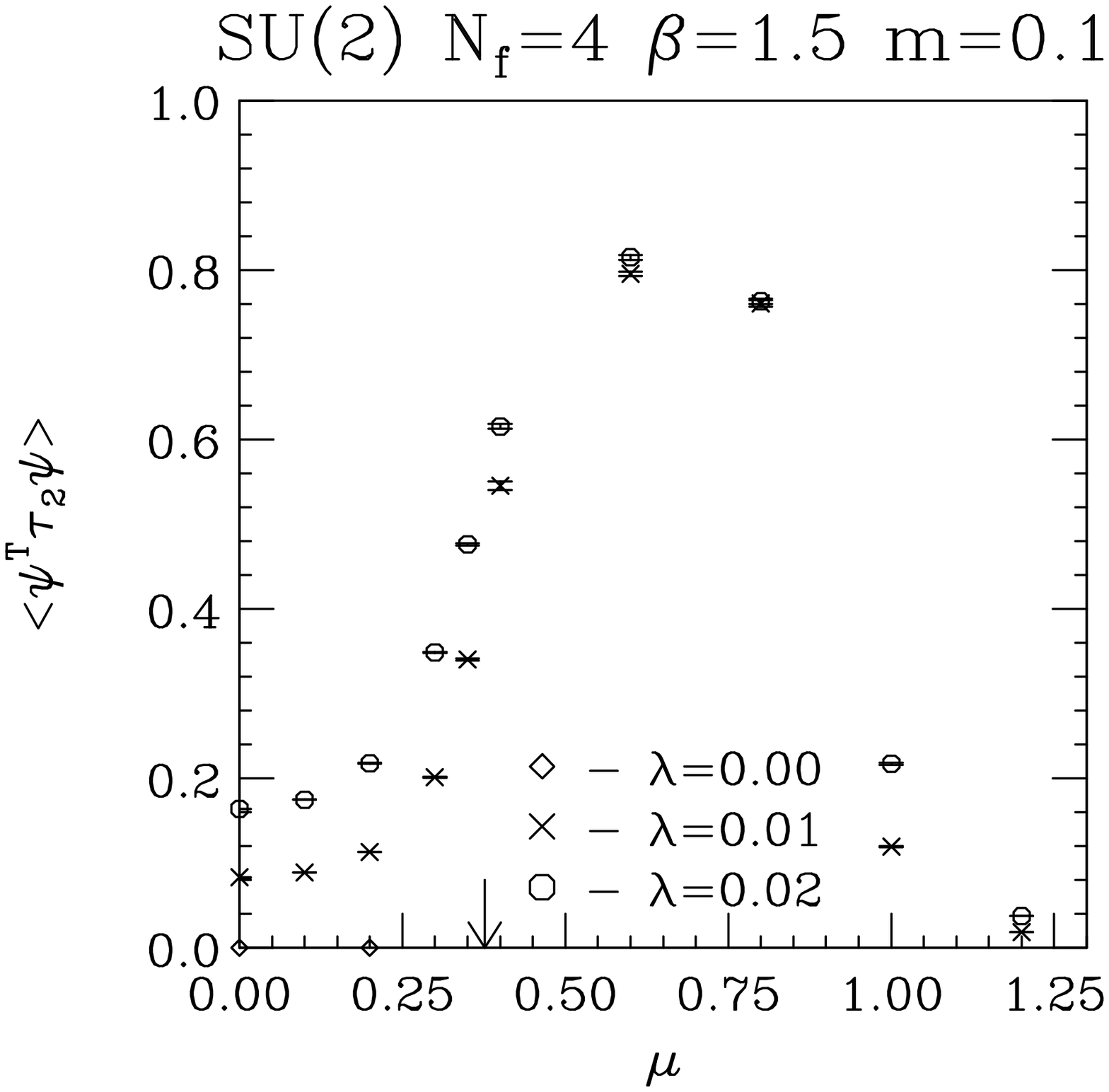}}}
\end{picture}
\caption{Chiral condensate (left) and diquark condensate (right) vs.\
chemical potential, for one flavour fundamental staggered quarks, on
an $8^4$ lattice; from
\protect\cite{Hands:2000hi}.  The arrow indicates $\mu\approx m_\pi/2$.}
\label{fig:sinclair-cond}
\end{figure}
\begin{figure}
\begin{picture}(11.5,6)
\put(2.5,0){\vbox{\epsfig{width=0.5\colw,file=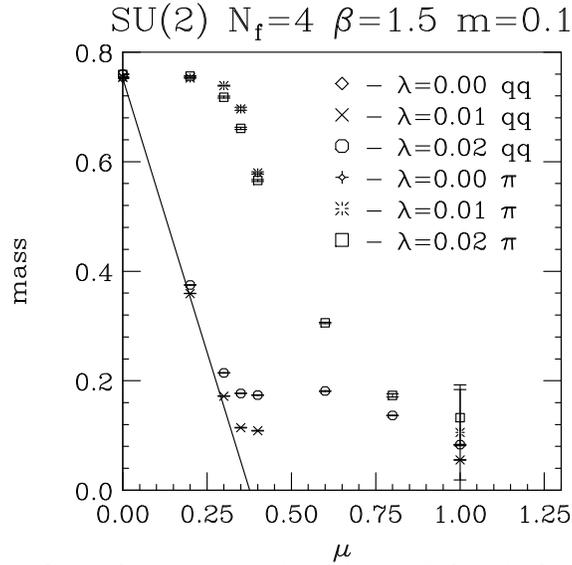}}}
\end{picture}
\caption{Pion and scalar diquark masses as functions of chemical
potential, for one flavour of fundamental staggered quarks, on an $8^4$
lattice; from \protect\cite{Hands:2000hi}.  The straight line is
$m=m_\pi-2\mu$.}
\label{fig:sinclair-mass}
\end{figure}


\begin{table*}[tbh]
\begin{tabular}{llllll}
 & \cc{$\mu$} & \multicolumn{3}{c}{TSMB} & \cc{HMC} \\ \cline{3-5}
 & & \cc{$\bra O\ket$} & \cc{$\bra O\ket_+$} & \cc{$\bra O\ket_-$} & 
\\ \hline
$\cbc$ & 0.0 & \m1.525(3) & & & \m1.526(1) \\
 & 0.36 & \m1.551(10) & 1.521(8) & 1.176(37) & \m1.485(9) \\
 & 0.4  & \m1.50(15) & 1.248(17) & 1.177(20) & \m1.253(10) \\ \cline{2-6}

$n$ & 0.0 & \m0.0000(13) & & & $-0.0002(3)$ \\
 & 0.36 & $-0.0003(80)$ & 0.0199(64) & 0.252(32) & \m0.0172(28) \\
 & 0.4  & \m0.07(10) & 0.177(14) & 0.208(15) & \m0.1667(90) \\
\end{tabular}
\caption{A comparison of results between TSMB and HMC for one flavour
of adjoint staggered quarks.  Due to long autocorrelation times, the
errors in the HMC results are probably
underestimated.\label{tab:tsmb}}
\end{table*}

\end{document}